\documentclass[english,aps,manuscript]{revtex4}
\usepackage{mathptmx}

\usepackage[T1]{fontenc}
\usepackage[latin9]{inputenc}
\usepackage{babel}
\usepackage{amsmath}
\usepackage{amssymb}
\usepackage{graphicx}
\usepackage{esint}
\usepackage[unicode=true,pdfusetitle,
 bookmarks=true,bookmarksnumbered=false,bookmarksopen=false,
 breaklinks=false,pdfborder={0 0 1},backref=false,colorlinks=false]
 {hyperref}

\makeatletter
\@ifundefined{textcolor}{}
{%
 \definecolor{BLACK}{gray}{0}
 \definecolor{WHITE}{gray}{1}
 \definecolor{RED}{rgb}{1,0,0}
 \definecolor{GREEN}{rgb}{0,1,0}
 \definecolor{BLUE}{rgb}{0,0,1}
 \definecolor{CYAN}{cmyk}{1,0,0,0}
 \definecolor{MAGENTA}{cmyk}{0,1,0,0}
 \definecolor{YELLOW}{cmyk}{0,0,1,0}
}

\makeatother

\begin{document}

\title{Ion-acoustic Shocks with Self-Regulated Ion Reflection and Acceleration }

\author{M.A. Malkov,$^{1}$ R.Z. Sagdeev,$^{2}$ G.I. Dudnikova,$^{2,4}$\\
T.V. Liseykina,$^{3,4}$ P.H. Diamond$,{}^{1}$ K. Papadopoulos,$^{2}$
C-S. Liu$,{}^{2}$ and J-J. Su$^{2}$}

\affiliation{$^{1}$CASS and Department of Physics, University of California,
San Diego, La Jolla, CA 92093\\$^{2}$University of Maryland, College
Park, MD 20742-3280\\$^{3}$Institut fuer Physik, Universitaet Rostock,
Rostock, Germany \\$^{4}$Institute of Computational Technologies
SD RAS, Novosibirsk, Russia }
\begin{abstract}
An analytic solution describing an ion-acoustic collisionless shock,
self-consistently with the evolution of shock-reflected ions, is obtained.
The solution extends the classic soliton solution beyond a critical
Mach number, where the soliton ceases to exist because of the upstream
ion reflection. The reflection transforms the soliton into a shock
with a trailing wave and a foot populated by the reflected ions. The
solution relates parameters of the entire shock structure, such as
the maximum and minimum of the potential in the trailing wave, the
height of the foot, as well as the shock Mach number, to the number
of reflected ions. This relation is resolvable for any given distribution
of the upstream ions. In this paper, we have resolved it for a simple
\textquotedblleft box\textquotedblright{} distribution. Two separate
models of electron interaction with the shock are considered. The
first model corresponds to the standard Boltzmannian electron distribution
in which case the critical shock Mach number only insignificantly
increases from $M\approx1.6$ (no ion reflection) to $M\approx1.8$
(substantial reflection). The second model corresponds to adiabatically
trapped electrons. They produce a stronger increase, from $M\approx3.1$
to $M\approx4.5$. The shock foot that is supported by the reflected
ions also accelerates them somewhat further. A self-similar foot expansion
into the upstream medium is also described analytically. 
\end{abstract}
\maketitle

\section{Introduction}

Collisionless shocks emerged in the 50s and 60s of the last century
as an important branch of plasma physics (see \cite{Sagdeev66,tidman1971shock,Kennel85,Papad85}
for review) and have remained ever since. Meanwhile, new applications
have posed new challenges to our understanding of collisionless shock
mechanisms. Particle acceleration in astrophysical settings, primarily
studied to test the hypothesis of cosmic ray origin in supernova remnant
shocks (see, e.g., \cite{BlandEich87,MDru01,BlandfordCRorig2014}
for review), stands out, and the collisionless shock mechanism is
the key. Among recent laboratory applications, a laser-based tabletop
proton accelerator is frequently highlighted as an affordable compact
alternative to the expensive synchrotron accelerators, currently used
to treat cancers \cite{Bulanov02,Haberberger2012NatPh,FiuzaUCLA2013}.

The goal of this article is twofold. First, we will obtain a self-consistent
analytic solution for the electrostatic structure of an ion-acoustic
collisionless shock with the Mach numbers beyond a critical value
$M=M_{*}\simeq1.6$ (for Boltzmannian electrons, and $M_{*}\approx3.1$
for adiabatically trapped electrons). At $M=M_{*},$ the shock is
about to reflect some of the upstream ions. Second, we will study
the dynamics of reflected ions, including their further acceleration.
A self-similar simple wave solution for electrostatic potential in
the foot region will be obtained selfconsistently with the incident
and reflected ion dynamics. We will show that an additional drop in
the foot electrostatic potential critically affects the ion reflection
from the main part of the shock. So, unlike most of the earlier analyzes,
treated the ion reflection using the test particle approximation,
e.g., \cite{Medvedev09,FiuzaUCLA2013}, we incorporate it into the
global shock structure. This study is relevant to the electrostatic
shock propagation in laser-produced plasmas, especially to the problem
of generation of monoenergetic ion beams, ion injection into the diffusive
shock acceleration in astrophysical shocks, and other shock-related
processes in astrophysical and space plasmas. 

In non-isothermal plasmas, with the electron temperature much higher
than ion temperature, $T_{e}\gg T_{i}$, a nonlinear Korteweg - de
Vries (KdV) equation applies as long as the nonlinearity remains \emph{weak}.
Of course, the KdV equation is famous for its soliton solution, one
of the most remarkable mathematical construction widely used in physics.
In plasmas, the solitons emerge when neither collisional nor Landau
damping is present. The \emph{ion-acoustic solitons}, in particular,
are the building blocks of collisionless shock waves at $T_{e}\gg T_{i}$.
Most lucidly they emerge from a solution pseudopotential, for an arbitrarily
\emph{strong} nonlinearity, thus comprising the limiting case of a
cnoidal wave solution with an infinite period \cite{Sagdeev66}. This
solution can also be interpreted as the uppermost ``energy level''
in a continuum of bound states in the pseudopotential, whereas the
lower energy bound states correspond to the periodic (cnoidal) waves.
The use of pseudopotential also illuminates formation of a soliton
wave-train, when even a small damping leads to the ``particle''
energy change in the pseudopotential which in reality corresponds
to the inner structure of the shock front \cite{MoisSagd63}. The
underlying mechanism here is the nonlinear Landau damping. Just a
few ions upstream reflected by the electric potential of the first
soliton will result in such damping. Then, by the \textquotedblleft nonlinear
saturation\textquotedblright{} effect, there are no more \textquotedblleft resonant\textquotedblright{}
ions to interact with the soliton train past the leading soliton. 

In the absence of resonant ions upstream, the first soliton breaks
down at $M>M_{*}\simeq1.6$ (this particular number is valid for cold
upstream ions and Boltzmann electrons). The solution ceases to exist
beyond this point, as there is no proper ``energy level'' in the
pseudopotential. This solution disappearance was thought to be the
point of \textquotedblleft overturning of the shock front\textquotedblright{}
and the end of the so-called \textquotedblleft laminar\textquotedblright{}
regime of ion-acoustic collisionless shocks. However, the results
of this paper prove otherwise. Namely, by including the reflected
ions into the shock structure, we have found the laminar solution
beyond $M=M_{*}$! More specifically, we found that when the ions
begin to reflect from the soliton tip at $M=M_{1}\lesssim M_{*}$,
the classical single soliton solution bifurcates into a more complex
structure. It comprises (i) the first soliton, (ii) the infinite periodic
wave train downstream of it, and (iii) the foot occupied by the reflected
ions. The front edge of the foot undergoes self-similar spreading
in a comoving reference frame of reflected ions. This solution continues
up to $M=M_{2}\gtrsim M_{*}$. 

At the second critical Mach number $M_{2}$, almost all incident ions
reflect, so the foot potential raises to increase the total shock
Mach number well above $M_{*}$. For the cold upstream ions, $T_{i}\ll T_{e}$,
$M_{1}$ approaches $M_{*}$, that is $M_{1}=M_{*}-\mathcal{O}\left(\sqrt{T_{i}/T_{e}}\right)$,
while $M_{2}\approx\sqrt{M_{*}^{2}+\left(1-1/4M_{*}^{2}\right)^{-1}\ln\left(1+\alpha\right)}$,
where $\alpha$ is the fraction of reflected ions. Note that $M_{2}\approx1.8$
for $\alpha=1$ and Boltzmannian electrons. The case of adiabatically
trapped electrons, in which $M_{*}\approx3.1$, gives a significantly
higher Mach number, $M_{2}\approx4.5$. The same pseudopotential technique
\cite{Sagdeev66}, also recovers the shock profile, although by introducing
two separate pseudopotentials $\Phi^{\pm}\left(\phi\right)=4\pi e\intop\left(n_{e}-n_{i}^{\pm}\right)d\phi$,
used for the plasma upstream and downstream of the leading soliton
($n_{i}^{+}\neq n_{i}^{-}$ due to the ion reflection). Here $\phi$
denotes the shock electrostatic potential. 

Within the range between the two critical points $M_{1}<M<M_{2}$,
the only time-dependent part of the solution is near the leading edge
of the reflected ion population. They support a pedestal upstream
of the leading soliton on which it rests. The reflected ions escape
upstream with double the shock speed in the pedestal reference frame,
Fig.\ref{fig:Electrostatic-potential-of}. Their further fate is determined
by a relatively slow spreading of the initially sharp front edge.
By even a small velocity dispersion, ions with higher initial velocity
undergo additional electrostatic acceleration by passing through the
shock pedestal. This process is described analytically as a self-similar
solution, which also yields the maximum velocity of reflected ions.

One usually employs two forms of electron density $n_{e}\left(\phi\right)$
in the pseudopotential. One form is the Boltzmannian, $n_{e}=n_{0}\exp\left(e\phi/T_{e}\right)$,
which yields $M_{*}\approx1.6$ \cite{Sagdeev66}. The other form
corresponds to adiabatically trapped electrons, in which case $M_{*}\approx3.1$
\cite{Gurevich68}. Depending on the practical situation either model
can be used. The Boltzmannian requires a Maxwellian distribution for
electrons trapped into potential wells (in analogy with barometric
formula). One can expect such scenario in the case when a higher density
plasma expands into a lower density (upstream) region. A suitable
example found in the conventional gas dynamics is a shock tube, in
which the shock is generated by breaking up a diaphragm, that was
separating the areas with different densities. By contrast, the adiabatic
trapping can be expected in a piston tube, in which the piston moves
into an initially uniform medium. Therefore, it models the shocks
generated in the pulsed laser-plasmas more accurately. Under these
circumstances, the production of reflected ions can be considered
as the laser-driven acceleration. It becomes more energy-efficient
at $M>M_{*}$, while producing almost monoenergetic ions over an extended
time interval.  

The paper is organized as follows. In Sec.\ref{sec:Shock-Model} we
discuss the shock model. Sec.\ref{sec:Solution-for-the} describes
the main part of the shock transition that forms in place of the parent
soliton after it has reflected a first few ions. Sec.\ref{sec:SolutionPrecursor}
presents a self-similar solution for the shock precursor supported
by reflected ions. We conclude with a Discussion in Sec.\ref{sec:Discussion}.

\section{The Shock Model\label{sec:Shock-Model}}

The analytic solution for an ion-acoustic soliton was first obtained
for the Boltzmannian electron distribution \cite{Sagdeev66} and extended
later to the case of adiabatically trapped electrons \cite{Gurevich68}.
Ions were assumed to be cold in both instances, which strictly limited
the maximum Mach numbers to $M_{*}\simeq1.6$ and $M_{*}\simeq3.1$
for the Boltzmann and adiabatic electrons, respectively. When the
Mach number reaches the maximum, the soliton begins to reflect some
of the upstream ions and the shock model must include them. Unlike
the soliton, the shock profile resulting from the ion reflection is
asymmetric about the reflection point. As shown in Ref.\cite{MoisSagd63},
its downstream part oscillates. Upstream of the soliton, reflected
ions will create a foot with an elevated electrostatic potential.

\begin{figure}
\includegraphics[bb=150bp 200bp 960bp 620bp]{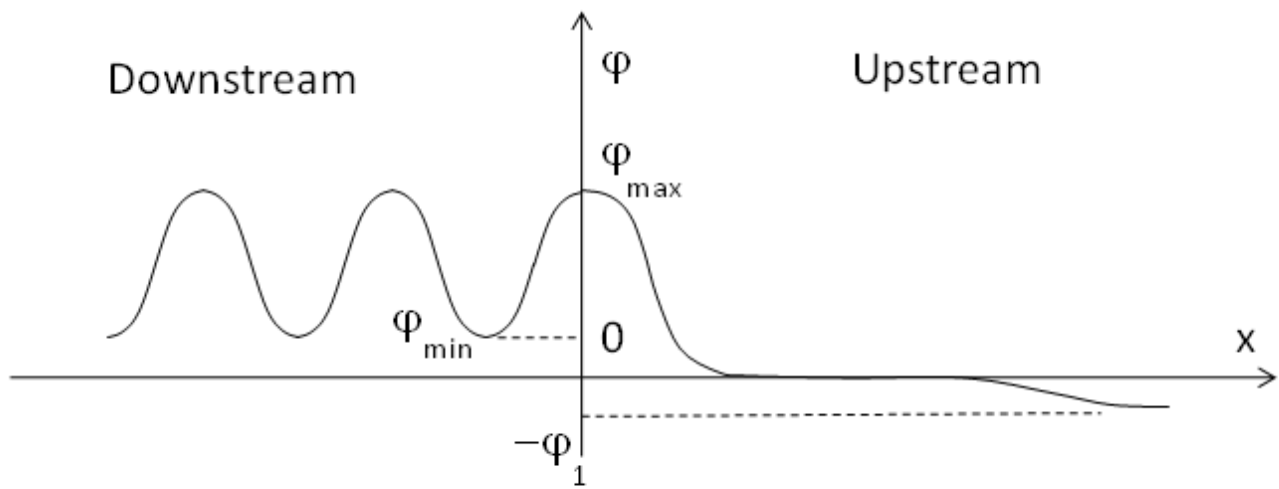}

\caption{Electrostatic potential of the shock structure consisting of a pedestal,
leading soliton and trailing wave\label{fig:Electrostatic-potential-of}}
\end{figure}

Seeking to extend the analytic solution beyond the ion reflection
point, we need a manageable reflection model. At a minimum, the model
should be able to relate the shock potential $\phi_{{\rm max}}$ and
Mach number $M$ to the number of reflected ions. Therefore, the model
must be kinetic, so one obtains the shock potential given the shock
speed and upstream ion distribution with a \emph{finite} temperature.
If the ion temperature upstream was zero ($V_{Ti}=0$) the ions would
reflect all at once when the shock Mach number crosses the point $M=$$\sqrt{2e\phi_{{\rm max}}/T_{e}}$.
By contrast, if $V_{Ti}\neq0$, then the reflection parameter $\alpha=n_{{\rm refl}}/n_{{\rm \infty}}$,
which is the ratio of reflected ion density to that of the incident
ions far away from the soliton, will continuously depend on the shock
parameters $M$ and $\phi_{{\rm max}}$. The region ahead of the shock
filled with the reflected ions of constant density (foot of the shock)
is mathematically regarded as ``infinity'' in the treatment of the
main part of the shock transition. There, all the relevant quantities,
such as the electrostatic potential $\phi$ are considered asymptotically
constant. The shock foot (precursor) will obviously expand linearly
with time after the first ions are reflected. In considering the main
part of the shock transition, we will count the plasma potential from
its value in the foot, so that we set the potential at ``infinity''
to $\phi=0$ in this section. Turning to the transition near the leading
edge of reflected ions in Sec.\ref{sec:SolutionPrecursor}, we will
account for the foot potential $\phi_{1}$ in the solution obtained
in this section, Fig.\ref{fig:Electrostatic-potential-of}.

To describe ion reflection we use a simple generalization of a cold
ion distribution upstream that provides an ion reflection model satisfying
the above requirements. So we use a ``box'' ion distribution with
the finite thermal velocity defined as $V_{Ti}=v_{2}-v_{1}$:

\begin{equation}
f_{i}^{\infty}\left(v\right)=\frac{1}{v_{2}-v_{1}}\begin{cases}
1, & -v_{2}<v<-v_{1}\\
0, & v\notin\left(-v_{2},-v_{1}\right)
\end{cases}\label{eq:BoxAppr}
\end{equation}
The normalization of $f_{i}^{\infty}$ implies a unity density of
incident ions far enough from the shock but not farther than the slowest
particles in the leading group of reflected ions at a given time,
as we discussed earlier. We use the shock frame throughout this section.
It is convenient to introduce a dimensionless potential by replacing
$e\phi/T_{e}\to\phi$ and measure the coordinate in units of $\lambda_{D}=\sqrt{T_{e}/4\pi e^{2}n_{\infty}}$,
while the ion velocity in units of the sound speed, $C_{s}=\sqrt{T_{e}/m_{i}}$. 

\begin{figure}
\includegraphics[bb=0bp 200bp 560bp 720bp]{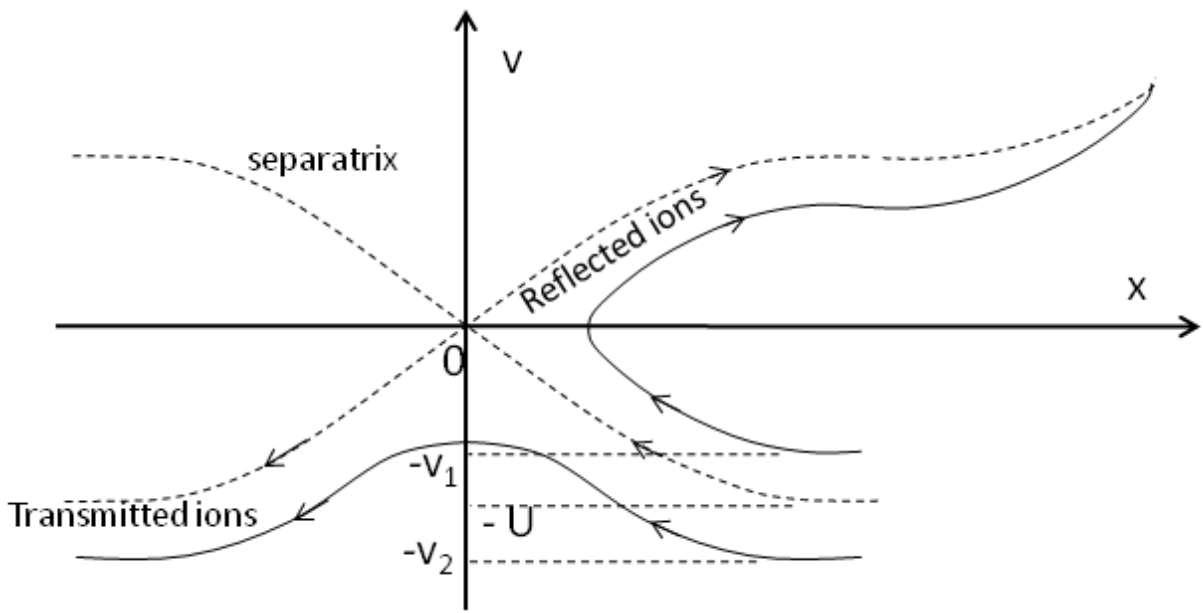}

\caption{Phase plane of ions at reflection point and propagation of reflected
ion beam accompanied by its further acceleration into the upstream
medium and narrowing its velocity distribution at large $x$. \label{fig:Phase-plane-of}}
\end{figure}

Suppose the soliton propagates in the positive $x$- direction with
a nominal speed $U=\sqrt{2\phi_{{\rm max}}}$ (w.r.t. the foot), where
$\phi_{{\rm max}}=\phi\left(0\right)$ is the maximum of its potential,
and $v_{1}\leq U\leq v_{2}$. The ion density upstream and downstream
can then be written as follows, Fig.\ref{fig:Phase-plane-of}

\begin{equation}
n_{i}\left(\phi\right)=\frac{1}{v_{2}-v_{1}}\begin{cases}
\sqrt{v_{2}^{2}-2\phi}-\sqrt{U^{2}-2\phi}, & x\leq0\\
\sqrt{v_{2}^{2}-2\phi}+\sqrt{U^{2}-2\phi}-2\sqrt{v_{1}^{2}-2\phi,} & x>0,\quad0<\phi<v_{1}^{2}/2\\
\sqrt{v_{2}^{2}-2\phi}+\sqrt{U^{2}-2\phi} & x>0,\quad v_{1}^{2}/2\leq\phi\leq U^{2}/2
\end{cases}\label{eq:niOfFi}
\end{equation}
Again, we count the electrostatic potential from its value in the
shock foot. We note that $U$ is not precisely the soliton velocity
but rather a convenient notation for $\sqrt{2\phi_{{\rm max}}}$,
while the soliton velocity with respect to the foot plasma (Mach number
in this reference frame) is $M=\left(v_{1}+v_{2}\right)/2$. The soliton
speed in the upstream plasma frame can only be determined when the
foot potential $\phi_{1}$ is obtained, Fig.\ref{fig:Electrostatic-potential-of}.
It is also important to note here that our choice of the simplest
form of ion distribution, eq.(\ref{eq:BoxAppr}) resulting in the
ion density including the reflected ions should lead to the same shock
structure in the limit $V_{Ti}\to0$ as in the case of, say, Maxwellian
distribution. In the latter case, the ion density in eq.(\ref{eq:niOfFi})
would be expressed through the error function. However, the limit
$V_{Ti}\to0$ can only be taken after the solution for the shock profile
is obtained.

From this point on, our treatment will depend on the particular electron
model, Boltzmannian or adiabatically trapped electrons. In the next
two subsections, these two models are considered separately.

\subsection{Boltzmannian Electrons}

Based on the above definitions, the Poisson equation for the shock
electrostatic potential can be written as follows

\begin{equation}
\frac{d^{2}\phi}{dx^{2}}=\left(1+\alpha\right)e^{\phi}-n_{i}\left(\phi\right)\label{eq:PoisGen}
\end{equation}
where 

\begin{equation}
\alpha=\frac{U-v_{1}}{v_{2}-v_{1}}\label{eq:Alpha}
\end{equation}
is the fraction of ions reflected off the shock, so that the first
term on the r.h.s of eq.(\ref{eq:PoisGen}) corresponds to the electron
contribution. We have chosen its normalization in such a way as to
neutralize the sum of the incident and reflected ions in the foot,
according to their normalization in eq.(\ref{eq:BoxAppr}). We may
now integrate eq.(\ref{eq:PoisGen}) once, also imposing the condition
$\phi^{\prime}\left(\phi_{{\rm max}}\right)=0$. The resulting equation
takes the following form

\begin{equation}
\frac{1}{2}\left(\frac{d\phi}{dx}\right)^{2}=\Phi\left(\phi\right)+\mathcal{F}^{\pm}\left(\phi\right)\equiv\Phi^{\pm}\left(\phi\right)\label{eq:PoisFirstIntPM}
\end{equation}
where $'+'$ or $'-$' sign should be taken for $x\geq0$ and $x<0$,
respectively. The functions $\Phi$ and $\mathcal{F}^{\pm}$ are given
by the following relations

\begin{equation}
\Phi=\left(1+\alpha\right)\left(e^{\phi}-e^{U^{2}/2}\right)+\frac{\left(v_{2}^{2}-2\phi\right)^{3/2}-\left(v_{2}^{2}-U^{2}\right)^{3/2}}{3\left(v_{2}-v_{1}\right)}\label{eq:FiDef}
\end{equation}
\begin{equation}
\mathcal{F}^{\pm}=\frac{1}{3\left(v_{2}-v_{1}\right)}\begin{cases}
\left(U^{2}-2\phi\right)^{3/2}-2\left(v_{1}^{2}-2\phi\right)^{3/2}\vartheta\left(v_{1}^{2}-2\phi\right), & x\geq0\\
-\left(U^{2}-2\phi\right)^{3/2}, & x<0
\end{cases}\label{eq:FpmDef}
\end{equation}
where $\vartheta$ is a Heaviside function. These relations are written
for the case $U\geq v_{1}$, while the opposite case would correspond
to the standard soliton solution with no ion reflection but with finite
upstream ion temperature, which we do not consider in this paper.
It is convenient to refer to the functions $\Phi^{\pm}\left(\phi\right)$
as to pseudopotentials of anharmonic oscillators of unit masses, whose
kinetic and potential energies correspond, respectively, to the l.h.s.
and the r.h.s of eq.(\ref{eq:PoisFirstIntPM}). Here, $\phi$ represents
the oscillator coordinate and $x$ represents time \cite{Sagdeev66}.
The shock structure $\phi\left(x\right)$ is thus completely determined
by eq.(\ref{eq:PoisFirstIntPM}) in a form of an inverse function
$x\left(\phi\right)$ under an appropriate choice of its branches
upstream and downstream. In the next section, we specify the critical
parameters of the shock profile $\phi_{{\rm max}}$ and $\phi_{{\rm min}}$,
depending on the upstream ion temperature and Mach number. Again,
by ``upstream'' we mean here the shock foot region where $\phi=0,$
Fig.\ref{fig:Electrostatic-potential-of}.

\subsection{Adiabatically Trapped Electrons\label{sub:Adiabatically-Trapped-Electrons-1}}

Boltzmann distribution for electrons near the shock that we considered
above is not always the best choice. If they drive the shock by themselves,
the shock may confine them, at least in part, to the downstream side
by trapping them in its potential. The trapped electrons acquire then
a flat distribution while the free electrons maintain their Maxwellian
distribution \cite{Gurevich68}. Hence, the following electron density
distribution replaces the Boltzmann distribution in the Poisson equation
given by eq.(\ref{eq:PoisGen}): 

\begin{equation}
\frac{d^{2}\phi}{dx^{2}}=\left(1+\alpha\right)\frac{F\left(\phi+\phi_{1}\right)}{F\left(\phi_{1}\right)}-n_{i}\left(\phi\right),\label{eq:PoisAdiab}
\end{equation}
where

\[
F\left(\phi\right)\equiv e^{\phi}{\rm erfc}\sqrt{\phi}+2\sqrt{\phi/\pi}
\]
while the ion distribution remains the same as in eqs.(\ref{eq:niOfFi},\ref{eq:PoisGen}).
In deriving eq.(\ref{eq:PoisAdiab}) we assumed electrons with negative
energy $E_{e}=mv^{2}/2-e\phi\left(x\right)\leq0$ to remain on the
downstream side of the shock structure, where their distribution $f_{0}$
is constant, while the rest of the electrons obey the standard Maxwell-Boltzmann
distribution. Apart from the normalization factor that accounts for
the ion reflection rate $\alpha$ and the finite shock foot potential
$\phi_{1}$, this distribution is identical to that used by Gurevich
in Ref. \cite{Gurevich68} for collisionless electrons trapped into
a soliton. Similarly to eq.(\ref{eq:PoisFirstIntPM}), the first integral
of the Poisson equation can be written as follows

\[
\frac{1}{2}\left(\frac{d\phi}{dx}\right)^{2}=\Phi_{a}\left(\phi\right)+\mathcal{F}^{\pm}\left(\phi\right)\equiv\Phi_{a}^{\pm}\left(\phi\right)
\]
where 

\begin{eqnarray}
\Phi_{a}\left(\phi\right) & = & \left(1+\alpha\right)\frac{F\left(\phi+\phi_{1}\right)-F\left(\phi_{{\rm max}}+\phi_{1}\right)-\frac{4}{3\sqrt{\pi}}\left[\left(\phi+\phi_{1}\right)^{3/2}-\left(\phi_{{\rm max}}+\phi_{1}\right)^{3/2}\right]}{F\left(\phi_{1}\right)}\label{eq:NeAdiab}\\
 & + & \frac{\left(v_{2}^{2}-2\phi\right)^{3/2}-\left(v_{2}^{2}-U^{2}\right)^{3/2}}{3\left(v_{2}-v_{1}\right)}
\end{eqnarray}
and $\mathcal{F}^{\pm}$ is given by eq.(\ref{eq:FpmDef}). Again,
we have added an integration constant to ensure that $d\phi/dx=0$
at $\phi=\phi_{{\rm max}}$. Once the shock model is defined for the
two types of electron distribution, we proceed with the solutions
for the respective shock structures.

\section{Solution for the main part of the shock transition\label{sec:Solution-for-the}}

\subsection{Boltzmannian Electrons}

An implicit solution for the potential $\phi\left(x\right)$ in the
regions $x\gtrless0$ may be written using eq.(\ref{eq:PoisFirstIntPM})
by the following inverse relations for $x\left(\phi\right)$

\begin{equation}
x\left(\phi\right)=\pm\frac{1}{\sqrt{2}}\intop_{\phi}^{\phi_{{\rm max}}}\frac{d\phi^{\prime}}{\sqrt{\Phi^{\pm}\left(\phi^{\prime}\right)}}\label{eq:xOFfi}
\end{equation}
At the point where ions are about to reflect off the soliton tip,
that is when $U=v_{1}$ ($\alpha=0$), two pseudopotentials are equal,
$\Phi^{+}=\Phi^{-}.$ Therefore, the (soliton) solution remains symmetric,
as it has to be in the case with no ion reflection. It is selected
by imposing an additional constraint on the pseudopotential $\Phi^{+}$.
Namely, $\Phi^{+}\left(\phi\right)$ must have a double root at $\phi=0$:
$\Phi^{+}\left(0\right)=0$, $\Phi^{+\prime}\left(0\right)=0$ regardless
of $\alpha$ being zero or positive. Note that the second condition,
$\Phi^{+\prime}\left(0\right)=0$ is satisfied automatically via our
choice of normalization of electron contribution, eq.(\ref{eq:PoisGen}),
that ensures charge neutrality at $+\infty$. The condition $d\phi/dx=0$
at $x=\infty$, that amounts to $\Phi^{+}\left(\phi=0\right)=0$,
yields the following nonlinear dispersion relation for the shock

\begin{equation}
\left(1+\alpha\right)\left(e^{U^{2}/2}-1\right)=\frac{v_{2}^{3}+U^{3}-2v_{1}^{3}-\left(v_{2}^{2}-U^{2}\right)^{3/2}}{3\left(v_{2}-v_{1}\right)}\label{eq:EqForU}
\end{equation}
Indeed, this is a relation between the shock amplitude $\phi_{{\rm max}}=U^{2}/2$
and its speed (Mach number w.r.t shock foot) $M=\left(v_{1}+v_{2}\right)/2$,
just as in the case of conventional ion-acoustic soliton of Ref.\cite{Sagdeev66}.
An important difference, however, is that this relation also includes
the ion reflection coefficient $\alpha=\left(U-v_{1}\right)/\left(v_{2}-v_{1}\right)$
and the upstream velocity dispersion $V_{Ti}=v_{2}-v_{1}$, through
which the upstream ion bounding velocities $v_{1}$ and $v_{2}$ in
eq.(\ref{eq:EqForU}) may always be expressed. In particular, $M=U+\left(1-2\alpha\right)V_{Ti}$.
Assuming that $V_{Ti}\ll U$, from eq.(\ref{eq:EqForU}) we obtain

\begin{equation}
e^{U^{2}/2}-1-U^{2}=-\frac{1}{3}\left(2U\right)^{3/2}\frac{\left(1-\alpha\right)^{3/2}}{1+\alpha}V_{Ti}^{1/2}\label{eq:EqForU-1}
\end{equation}
The left hand side (l.h.s.) of this relation is identical to the soliton
dispersion relation (l.h.s.=0) taken at the ion reflection potential
($\phi_{{\rm max}}=U^{2}/2$). Therefore, the ion reflection does
not change the shock speed w.r.t. the shock precursor in a plasma
with cold ions upstream, $V_{Ti}\to0$. Comparing eqs.(\ref{eq:EqForU})
with (\ref{eq:EqForU-1}) we see how this results from canceling out
of the factor $1+\alpha$. However, the shock speed does grow with
ion reflection rate $\alpha$ w.r.t. the upstream frame (since the
precursor height $\phi_{1}$ grows as well), which we discuss later.
It is also interesting to observe that the thermal correction to the
shock speed diminishes with an increase in ion reflection, $\alpha\to1$.

Neglecting the r.h.s. (cold upstream ions, $V_{Ti}\to0$, or $\alpha\to0$)
gives the solution for the critical Mach number $U=M_{*}\approx1.6$
\cite{Sagdeev66}. For a finite $V_{Ti}\ll v_{1,2}$ and arbitrary
$\alpha<\alpha_{c}\approx1$ (see below), we obtain the following
dispersion relation

\begin{equation}
U\approx M_{*}-\frac{2^{3/2}}{3}\frac{\left(1-\alpha\right)^{3/2}M_{*}^{1/2}V_{Ti}^{1/2}}{\left(1+\alpha\right)\left(M_{*}^{2}-1\right)}\label{eq:DispRelWithRefl}
\end{equation}

\begin{figure}
\includegraphics[bb=0bp 200bp 560bp 720bp,clip]{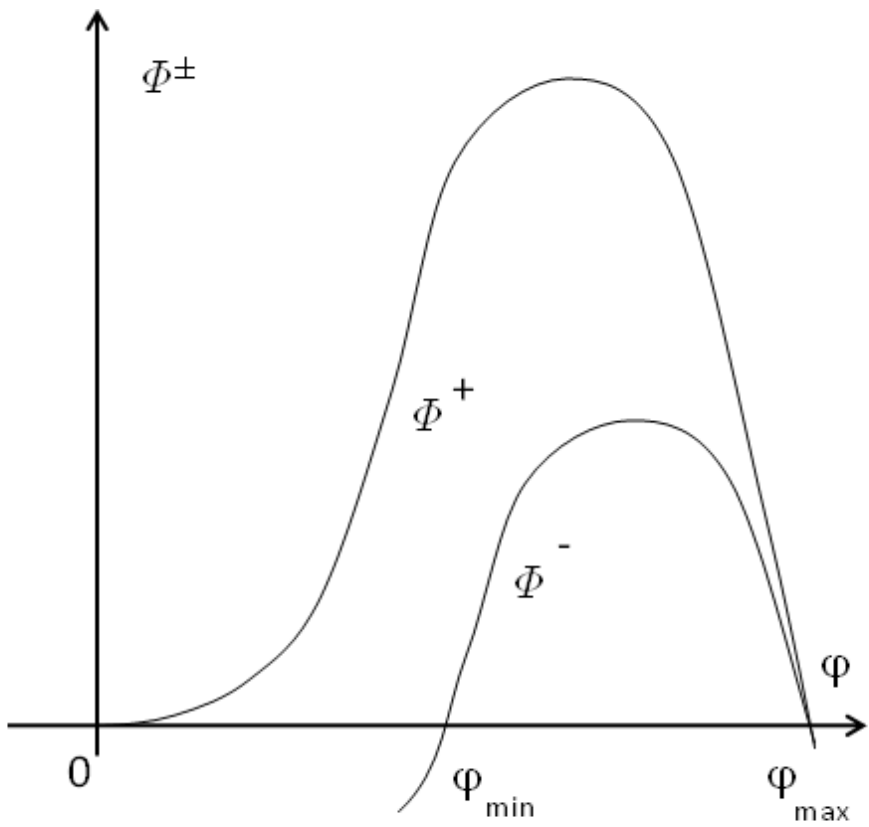}

\caption{Pseudopotentials of ``oscillators'' described by eq.(\ref{eq:PoisFirstIntPM}).\label{fig:Pseudopotentials-of-oscillators}}
\end{figure}

Turning to the spatial profile of the potential downstream ($x<0$),
from eq.(\ref{eq:PoisFirstIntPM}) and Fig.\ref{fig:Pseudopotentials-of-oscillators}
we see that it oscillates between its minimum value $\phi_{{\rm min}}$
and $\phi_{{\rm max}}=U^{2}/2$ that is given by eq.(\ref{eq:DispRelWithRefl}).
Similarly to the above equation for $\phi_{{\rm max}}$, given by
eq.(\ref{eq:EqForU-1}), from eq.(\ref{eq:PoisFirstIntPM}) we obtain
the following equation for $\phi_{{\rm min}}$

\begin{equation}
e^{U^{2}/2}-e^{\phi_{{\rm min}}}=\frac{\left(v_{2}^{2}-2\phi_{{\rm min}}\right)^{3/2}-\left(U^{2}-2\phi_{{\rm min}}\right)^{3/2}-\left(v_{2}^{2}-U^{2}\right)^{3/2}}{3\left(1+\alpha\right)\left(v_{2}-v_{1}\right)}\label{eq:EqForFimin}
\end{equation}
The solution for $\phi_{{\rm min}}$ simplifies for the cases of weak
and strong reflection. So, for $\alpha\ll1$, using also eq.(\ref{eq:DispRelWithRefl})
we find

\[
\phi_{{\rm min}}\simeq\frac{2M_{*}^{2}\sqrt{\alpha}}{\sqrt{M_{*}^{2}-1}}
\]
The opposite case of strong reflection, $1-\alpha\ll1$, should be
treated with care when the small parameter $1-\alpha$ approaches
the thermal spread of incident ions, $V_{Ti}$. First, assuming that
$V_{Ti}\ll1-\alpha\ll1$, we obtain

\[
\phi_{min}\simeq\phi_{max}-\frac{2M_{*}^{2}\left(1-\alpha\right)^{2}}{\left(1+M_{*}^{2}\right)^{2}}
\]
For smaller $1-\alpha$ we may write

\begin{equation}
\phi_{min}\simeq\phi_{max}-\frac{9}{4}\left(1-\alpha\right)V_{T_{i}}M_{*}\left[1-\sqrt{\frac{V_{Ti}}{2\left(1-\alpha\right)}}\frac{\left(1+M_{*}^{2}\right)}{\sqrt{M_{*}}}\right]^{2}\label{eq:FiMin2}
\end{equation}
The last solution cannot be continued to $\alpha=1$, as $\phi_{{\rm min}}$
reaches $\phi_{{\rm max}}$ at 
\begin{equation}
\alpha=\alpha_{c}\simeq1-V_{Ti}\left(1+M_{*}^{2}\right)^{2}/2M_{*}<1\label{eq:AlphaC}
\end{equation}
It is not difficult to understand why there is no solution corresponding
to complete ion reflection as $\alpha_{c}\neq1$. Indeed, a solution
with all particles reflected from the shock would nevertheless require
a finite density downstream (to neutralize electrons), which could
be possible only if the incident ions had no velocity dispersion (that
is why $1-\alpha_{c}\sim V_{Ti}$ in eq.{[}\ref{eq:AlphaC}{]}). Therefore,
when $\alpha$ increases to $\alpha=\alpha_{c}$, a solution $\phi\left(x\right)\equiv\phi_{{\rm max}}=const$
establishes downstream (pure shock transition). Instead of using eqs.(\ref{eq:EqForU})
and (\ref{eq:FiMin2}), this special solution is easier to find directly
by requiring charge neutrality condition fulfilled identically downstream,
eq.(\ref{eq:PoisGen})

\[
e^{U^{2}/2}=\frac{1-\alpha}{1+\alpha}\sqrt{\frac{2U}{V_{Ti}\left(1-\alpha\right)}+1}.
\]
This result, in combination with eq.(\ref{eq:EqForU-1}), yields the
critical value $\alpha=\alpha_{c}$ in eq.(\ref{eq:AlphaC}). Under
this condition, the maximum potential $\phi_{{\rm max}}=U^{2}/2$
is determined by 

\begin{equation}
U=M_{*}-\frac{V_{Ti}^{2}}{3M_{*}^{2}}\frac{\left(M_{*}^{2}+1\right)^{3}}{M_{*}^{2}-1}.\label{eq:Uupper}
\end{equation}
Together with eq.(\ref{eq:DispRelWithRefl}), the latter expression
constrains the range of the shock Mach numbers for $0<\alpha<\alpha_{c}\approx1$.
These values of the shock potential and Mach number ($\simeq U$)
relate to the upstream region occupied by reflected ions (where $\phi=0$).
This region is located to the right from the leading soliton, but
not farther than the slowest ions out of those that have been reflected
first, Fig.\ref{fig:Electrostatic-potential-of}. A more precise meaning
of this condition will be given in the next section. Now we turn to
the calculation of the shock parameters for adiabatically trapped
electrons.

\subsection{Adiabatically Trapped Electrons\label{sub:Adiabatically-Trapped-ElectronsSolution}}

The calculation of shock characteristics for adiabatically trapped
electrons is similar to that for the Boltzmannian electrons, but with
one significant difference: the foot potential $\phi_{1}$ explicitly
enters the Poisson equation also for the main part of the shock transitioin,
cf. eqs.(\ref{eq:PoisGen},\ref{eq:PoisAdiab}). Therefore, unlike
in the Boltzmannian case, where the ion reflection rate $\alpha$
explicitly enters the shock solution  only in conjunction with the
incident ion thermal spread (see, e.g., eq.{[}\ref{eq:DispRelWithRefl}{]}),
in the case of adiabatically trapped electrons the ion reflection
effect is significantly stronger. The foot elevation $\phi_{1}$ is
the largest contributing factor to that. Although, the latter is also
determined by $\alpha$ that we will discuss in the next section.

\begin{figure}
\includegraphics[scale=0.7,angle=270]{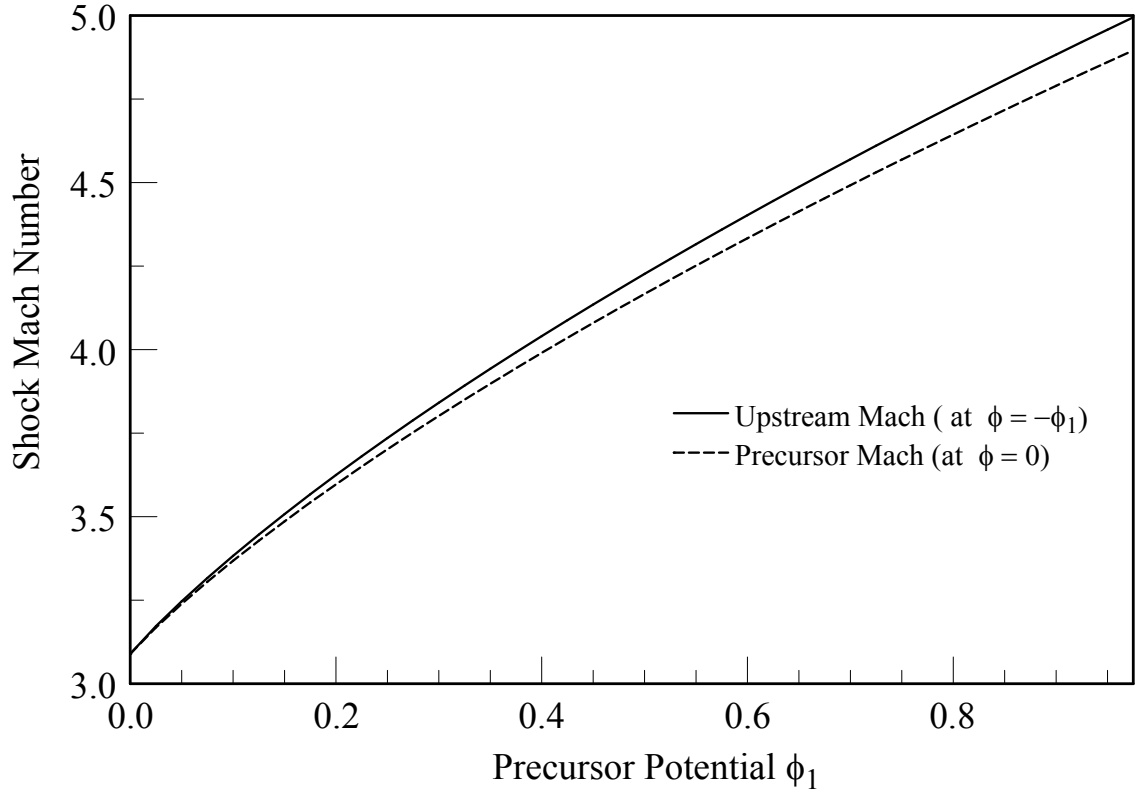}

\caption{Solution of eq.(\ref{eq:EqForFimAd}) in the limit $V_{Ti}=0$ shown
in the form of the shock Mach number related to the upstream frame,
$M$, and to the foot ion reference frames, $U=\sqrt{2\phi_{{\rm max}}}$
(see text).\label{fig:AdMach}}

\end{figure}

Turning now to the shock solution, for its potential $\phi_{{\rm max}}\equiv U^{2}/2$
we obtain from eq.(\ref{eq:NeAdiab}) the following relation:

\begin{equation}
\frac{F\left(\phi_{{\rm max}}+\phi_{1}\right)+\frac{4}{3\sqrt{\pi}}\left[\left(\phi_{{\rm max}}+\phi_{1}\right)^{3/2}-\phi_{1}^{3/2}\right]}{F\left(\phi_{1}\right)}-1-2\phi_{{\rm max}}=-\frac{2^{9/4}}{3}\phi_{{\rm max}}^{3/4}\frac{\left(1-\alpha\right)^{3/2}}{1+\alpha}V_{Ti}^{1/2}\label{eq:EqForFimAd}
\end{equation}
Unlike in the Boltzmann case, where the shock amplitude $\phi_{{\rm max}}$
was given by just a number $M_{*}^{2}/2$ in the limit $V_{Ti}\to0$,
now $\phi_{{\rm max}}$ depends directly on $\phi_{1}$. This dependence
can be determined by solving eq.(\ref{eq:EqForFimAd}) numerically
for $V_{Ti}\to0$, as the major contributing factor is $\phi_{1}$.
Fig.\ref{fig:AdMach} shows this solution in the form of the shock
Mach number related to the far upstream medium and to the shock precursor,
where $\phi=0$. The latter is given by the relation $U\left(\phi_{1}\right)=\sqrt{2\phi_{{\rm max}}}$,
shown with the dashed line. As expected, it starts from the value
$U\approx3.1$, which is the maximum speed of a non-reflecting soliton
calculated by Gurevich \cite{Gurevich68}. As the ion reflection rate
$\alpha$ increases, so do $\phi_{1}\left(\alpha\right)$ and $U$.

To calculate the shock Mach number in the far upstream reference frame
rather than the foot frame, one has to take into account the foot
potential $\phi_{1}$. Indeed, the incident ions first slow down by
passing through the potential $\phi_{1}$ before they hit the leading
soliton in the shock structure and specularly reflect off it. The
total Mach number (i.e. the absolute shock speed, again, given for
$V_{Ti}\to0$) then amounts to

\begin{equation}
M=2^{3/2}\sqrt{\phi_{{\rm max}}+\phi_{1}/4}-\sqrt{2\phi_{{\rm max}}}\approx\sqrt{2\phi_{{\rm max}}+\phi_{1},}\label{eq:MachTot}
\end{equation}
where the last expression is an approximation for $\phi_{1}\ll2\phi_{{\rm max}}$
which holds up reasonably well for even a strong ion reflection. This
dependence is shown in Fig.\ref{fig:AdMach} with the solid line.
We see from the last equation that the direct effect of the foot elevation
$\phi_{1}$ on the total Mach number is quite small, so that in the
case of Boltzmannian electrons, where $\phi_{{\rm max}}$ does not
depend on $\phi_{1}$ explicitly (eq.{[}\ref{eq:EqForU}{]}) the maximum
Mach number remains close to $M_{*}$. The dependence of the Mach
number on $\alpha$ is considerably stronger for adiabatically trapped
electrons, where $\phi_{{\rm max}}$ explicitly depends on $\phi_{1}\left(\alpha\right)$.

Now that we have obtained the shock structure up the location of first
reflected ions, we turn to the dynamics of these ions. Obviously,
they are at the front end of the shock precursor, where the shock
potential drops to its upstream value. Depending on the time elapsed
from the moment when the first ion reflection occurred, this area
may be far away from the main part of the entire shock structure.
Therefore, the assumption about the adiabatic trapping of electrons
may not be justified there, so we restrict our calculation of $\phi_{1}$
to the case of Boltzmannian electrons. In the next section, we will
relate the foot potential $\phi_{1}$ to $\alpha$. This relation
provides the shock parameters depending only on the reflection parameter
$\alpha$, by using eq.(\ref{eq:MachTot}) and Fig.\ref{fig:AdMach}.

\section{Solution for the ion precursor\label{sec:SolutionPrecursor}}

As we have seen, in the case of $\alpha>0$ the shock propagates through
a foot region with the electrostatic potential elevated to $\psi=\phi_{1}$
from its level $\psi=0$ at $+\infty$. By entering this area the
incident ions slow down before they encounter the leading soliton.
It is convenient to account for this change in the shock structure
potential by shifting the variable $\phi$ used in the previous sections
to $\psi$ as follows $\psi=\phi+\phi_{1}$. So, we now focus on that
part of the shock transition where $\psi$ varies in the interval
$0<\psi<\phi_{1}$, Fig.\ref{fig:Electrostatic-potential-of}. From
the physics perspective, one may assume that after an initial propagation
of the reflected beam upstream, a quasi-steady flow at a constant
speed and potential $\psi=\phi_{1}$ will be established between the
ion-reflecting soliton and the head of the beam. Our consideration
of the beam dynamics below (see, in particular, Appendix) implies
that this assumption is not justified if the beam reflection is not
strictly stationary, so the beam may continue to evolve all the way
through its extension. However, under a stationary reflection most
of the beam will also be stationary, and only its head will continue
to spread via a self-induced electric field. In this region, the potential
will gradually decrease from $\psi=\phi_{1}$ to $\psi=0$, and the
respective electric field will further accelerate the reflected particles.
We will also see that although the ion velocity distribution narrows
locally in the course of acceleration, the spatially averaged distribution,
in fact, broadens. Therefore, if the reflected beam impinges on a
target upstream, for example, the average particle energy will be
decreasing until the target has absorbed all the transient part of
the beam coming from the region where $0<\psi<\phi_{1}$. Then, the
stationary part of the reflected beam, which carries potential $\phi_{1}$
reaches the target. This phase, characterized by the constant energy
deposition, will continue until the leading edge of the shock arrives
at the target. 

In describing the pedestal part of the shock transition, it is natural
to use the reference frame in which the stationary part of the reflected
ion beam is at rest. It is also clear that this part of the shock
profile can be described almost independently of the main part of
the shock transition, presented earlier in the paper. Then, a matching
condition in the region where $\phi=0$ or, equivalently, $\psi=\phi_{1}$,
applies. Here, all the relevant particle groups have a constant density.
Using the plasma neutrality requirement for the Boltzmannian electrons
and ions that enter this region at the speed -w from $+\infty$, we
obtain

\begin{equation}
e^{\phi_{1}}=\frac{1+\alpha}{\sqrt{1-2\phi_{1}/w^{2}}}\label{eq:fi1Def}
\end{equation}
which, for $2\phi_{1}/w^{2}\ll1$, can be written also as $\exp\left[\left(1-w^{-2}\right)\phi_{1}\right]\approx1+\alpha$.
Hence, 

\begin{equation}
\phi_{1}\simeq\frac{\ln\left(1+\alpha\right)}{1-w^{-2}}\label{eq:fi1}
\end{equation}
for any $0<\alpha<1$. In the current reference frame, the velocity
of the incoming upstream ions is $-w\simeq-2U$, so the requirement
for eq.(\ref{eq:fi1}) to be valid is $2\phi_{1}/w^{2}\approx\phi_{1}/2\phi_{{\rm \max}}\ll1$.
This condition is warranted by eq.(\ref{eq:fi1}), if $\alpha\ll1$
but, because $\phi_{1}$ is not large even for $\alpha\lesssim1$,
the approximate formula in eq.(\ref{eq:fi1}) is accurate to within
$1\%$ for all $0<\alpha<1$. The thermal spread of ions is neglected
here. To further simplify notations, we rescale the spatial variable
$x$ here as follows $x^{\prime}=\left(1-2\phi_{1}/w^{2}\right)^{1/4}x$.
This change of variable is not significant for the sequel, though.
Denoting the reflected ion density by $\rho$ we can write the Poisson
equation in the following way

\begin{equation}
\frac{d^{2}\psi}{dx^{\prime2}}=e^{\psi}-\frac{1}{\sqrt{1-2\psi/w^{2}}}-\rho.\label{eq:PoisPrec}
\end{equation}
The limiting values for $\rho$ are $\rho\left(\psi=0\right)=0$ and 

\begin{equation}
\rho\left(\psi=\phi_{1}\right)=\frac{\alpha}{\sqrt{1-2\phi_{1}/w^{2}}}.
\end{equation}
As before, we assume that the front-running beam particles already
escaped the main part of the shock and have spread to an area much
larger than the Debye length. Hence, the following ``quasi-neutral''
version of eq.(\ref{eq:PoisPrec}) applies 

\begin{equation}
\rho=e^{\psi}-\frac{1}{\sqrt{1-2\psi/w^{2}}}.\label{eq:EOS}
\end{equation}
The last relation can be used as an equation of state of the reflected
ion gas. Furthermore, at this stage the problem of subsequent spreading
of the reflected beam lacks any characteristic length. Therefore,
as in the case of its gasdynamics counterpart, the solution should
depend only on the variable 

\[
\xi=x/t.
\]
Placing the spreading front edge of the reflected ion beam at the
origin, we obtain the following boundary conditions for the beam density,
its velocity and plasma potential: $\rho\left(\infty\right)=u(-\infty)=\psi\left(\infty\right)=0$,
$\psi\left(-\infty\right)=\phi_{1}$, and

\begin{equation}
\rho\left(-\infty\right)=\frac{\alpha}{\sqrt{1-2\phi_{1}/w^{2}}}\equiv\rho_{1}.\label{eq:rho1}
\end{equation}
An additional limitation to this treatment, that uses the particle
energy conservation in eqs.(\ref{eq:fi1Def}-\ref{eq:PoisPrec}),
is that the potential $\psi$ should not vary significantly during
the crossing time of incident ions. The reason for such variation
is, of course, the spreading of the reflected ions entering eq.(\ref{eq:PoisPrec})
through the term $\rho\left(x,t\right)$. Again, the above limitation
is easily fulfilled as $\phi_{1}\ll w^{2}/2$ even for $\alpha\sim1$.
By neglecting also the ion pressure, we arrive at the following hydrodynamic
equations for the reflected ions:

\begin{equation}
\frac{\partial\rho}{\partial t}+\frac{\partial}{\partial x}\rho u=0\label{eq:ContPrec}
\end{equation}

\begin{equation}
\frac{\partial u}{\partial t}+u\frac{\partial u}{\partial x}=-\frac{\partial\psi}{\partial x},\label{eq:MomenPrec}
\end{equation}
where $u$ is the flow velocity of reflected ions in the comoving
reference frame. As we use dimensionless variables introduced in Sec.
\ref{sec:Shock-Model} for $x$, $u$ and $\psi$, time is now measured
in the units of $\omega_{pi}^{-1}=\sqrt{m_{i}/4\pi e^{2}n_{\infty}}$. 

The problem, given by eqs.(\ref{eq:PoisPrec},\ref{eq:ContPrec}-\ref{eq:MomenPrec}),
has a close relation to the problem of expansion of one gas into another
(or into vacuum) \cite{LLFM,GurevichPit75}. Indeed, as we assume
the pedestal having already spread to a region larger than the Debye
length, we use quasi-neutrality condition, eq.(\ref{eq:EOS}) in place
of the Poisson equation (\ref{eq:PoisPrec}). This implies $\psi=\psi\left(\rho\right)$
(simple wave solution) and the r.h.s. of eq.(\ref{eq:MomenPrec})
corresponds to the specific enthalpy gradient of the gasdynamics analog
of eqs.(\ref{eq:ContPrec}-\ref{eq:MomenPrec}). Looking for such
solution, from eqs.(\ref{eq:ContPrec}-\ref{eq:MomenPrec}) we obtain
(see also Appendix for a more general treatment of eqs.{[}\ref{eq:ContPrec}-\ref{eq:MomenPrec}{]})

\begin{equation}
\left[\left(u-\xi\right)^{2}-\left(\frac{\partial\ln\rho}{\partial\psi}\right)^{-1}\right]\frac{\partial\psi}{\partial\xi}=0,\label{eq:SimpleWaveEq}
\end{equation}
where

\begin{equation}
u=\intop_{\psi}^{\phi_{1}}d\psi\sqrt{\frac{\partial\ln\rho}{\partial\psi}}\label{eq:uOfPsi}
\end{equation}
and $\rho\left(\psi\right)$, again, obeys the ``equation of state''
of the reflected ion gas given by eq.(\ref{eq:EOS}). Eq.(\ref{eq:SimpleWaveEq}),
in turn, is satisfied by the following piecewise continuous solution

\begin{equation}
\psi=\begin{cases}
\phi_{1}, & \xi\le\xi_{1}<0\\
0, & \xi\ge\xi_{2}>0
\end{cases}\label{eq:PsiSol}
\end{equation}
In the expanding wave region $\xi_{1}<\xi<\xi_{2}$, the solution
is given by

\begin{equation}
u=\xi+\left(\frac{\partial\ln\rho}{\partial\psi}\right)^{-1/2}\label{eq:uSimpleWaveSol}
\end{equation}
Together with eq.(\ref{eq:uOfPsi}), the last equation determines
the profile of the expanding wave in the form of $\xi\left(\psi\right)$:

\begin{equation}
\xi\left(\psi\right)=\intop_{\psi}^{\phi_{1}}d\psi\sqrt{\frac{\partial\ln\rho}{\partial\psi}}-\left(\frac{\partial\ln\rho}{\partial\psi}\right)^{-1/2}\label{eq:ksiOfPsi}
\end{equation}
By applying the boundary conditions $\psi\left(\xi_{1}\right)=\phi_{1}$
and $\psi\left(\xi_{2}\right)=0$, for the edges $\xi_{1,2}$ of the
simple wave, given by eq.(\ref{eq:PsiSol}-\ref{eq:uSimpleWaveSol}),
we obtain 

\begin{equation}
\xi_{1}=-\left(\frac{\partial\ln\rho}{\partial\psi}\right)_{\psi=\phi_{1}}^{-1/2}\label{eq:ksi1}
\end{equation}

\begin{equation}
\xi_{2}=\intop_{0}^{\phi_{1}}d\psi\sqrt{\frac{\partial\ln\rho}{\partial\psi}}.\label{eq:ksi2}
\end{equation}
These are the velocities with which the simple wave expands back into
the beam and the upstream plasma, respectively. 

From eq.(\ref{eq:uOfPsi}), for the maximum beam velocity (at $\psi=0$)
we obtain $u_{{\rm max}}\simeq2\sqrt{\phi_{1}}$. The total speed
of the shock is $M\simeq\sqrt{2\phi_{\max}+\phi_{1}}$ ($\phi_{1}\ll2\phi_{{\rm max}}$,
eq.{[}\ref{eq:MachTot}{]}). Neglecting the upstream ion temperature
in eqs.(\ref{eq:DispRelWithRefl}-\ref{eq:Uupper}) and using eq.(\ref{eq:fi1}),
this Mach number can be written as $M\approx\sqrt{M_{*}^{2}+\left(1-1/4M_{*}^{2}\right)^{-1}\ln\left(1+\alpha\right)}$,
which yields $M=M_{{\rm max}}\approx1.8$ for $\alpha\approx1$ under
a Boltzmannian electron distribution. The maximum reflected beam speed
w.r.t. the upstream rest frame is $V_{b}=M+M_{*}+u_{{\rm max}}\approx2\left[M_{*}+\sqrt{\ln\left(1+\alpha\right)}\right]$.
For the adiabatically trapped electrons, the calculation of $M\left(\alpha\right)$
is somewhat more complicated since the shock maximum potential $\phi_{{\rm max}}$
explicitly depends on $\phi_{1}$, as we discussed in Sec.\ref{sub:Adiabatically-Trapped-ElectronsSolution}.

\subsection{Acceleration of reflected ions}

It follows that, even when ions are bouncing off the shock front,
the laminar shock structure persists for up to a maximum Mach number
$M_{{\rm max}}$. This value is somewhat higher than the classical
limit $M=M_{*}\approx1.6$ for the Boltzmannian electrons ($M_{{\rm max}}\approx1.8$)
and considerably higher for adiabatically trapped electrons, where
$M_{*}\approx3.1$, Fig.\ref{fig:AdMach}. In the meanwhile, the fraction
of reflected particles may approach almost unity, eq.(\ref{eq:AlphaC}).
At $\psi=0$ in eq.(\ref{eq:uOfPsi}), the reflected beam velocity
reaches its maximum. By expanding eq.(\ref{eq:uOfPsi}) for small
$\alpha$ we obtain $u_{{\rm max}}\simeq2\sqrt{\phi_{1}}$ which is
a factor of $\sqrt{2}$ higher than what the front running particles
would gain from the energy conservation after being accelerated from
the shock foot of a height $\phi_{1}$. The difference is explained
by the expansion of reflected particles. 

An equally important aspect of the reflected beam dynamics is that
the beam, while being accelerated by the self-generated electric field,
substantially narrows its velocity distribution. Indeed, consider
the beam temperature evolution during its expansion upstream. As before,
we neglect the internal pressure of the beam in the hydrodynamic equations
(\ref{eq:ContPrec}) and (\ref{eq:MomenPrec}) that describe the flow.
But once we have described the ion beam flow, we may also calculate
the evolution of its temperature in a test-particle regime. Assuming
that the beam expands adiabatically, the equation for its temperature
$T_{b}$ takes the following form 

\begin{equation}
\frac{\partial T_{b}}{\partial t}+u\frac{\partial T_{b}}{\partial x}+\left(\gamma-1\right)T_{b}\frac{\partial u}{\partial x}=0\label{eq:TbAdiab}
\end{equation}
where $\gamma$ is the ion adiabatic index. By combining this equation
with the continuity eq.(\ref{eq:ContPrec}), we obtain

\begin{equation}
\frac{T_{b}\left(\psi\right)}{T_{b}\left(\phi_{1}\right)}=\left[\frac{\rho\left(\psi\right)}{\rho\left(\phi_{1}\right)}\right]^{\gamma-1}\label{eq:BeamCool}
\end{equation}
where $T_{b}\left(\phi_{1}\right)$ is the reflected beam temperature
in the foot region where $\psi=\phi_{1}$, Fig.\ref{fig:Electrostatic-potential-of}.
For the simple ``box'' model, $T_{b}\left(\phi_{1}\right)=\left(v_{2}-U\right)^{2}/24$.
The result shown in eq.(\ref{eq:BeamCool}) is, as expected, just
a familiar adiabatic law. Asymptotically, the width of reflected ion
beam distribution narrows down to zero far upstream where $\psi\to0$.
Note that the local beam density also vanishes ($\rho\to0$) at this
point, according to eq.(\ref{eq:EOS}). We see from eq.(\ref{eq:BeamCool})
that the most efficient energy collimation occurs in 1D motion ($\gamma=3$),
e.g., if there is a strong magnetic field present. 

Unfortunately, the beam energy changes in space (and time) while it
accelerates through the pedestal region, where the potential $\psi$
changes between $0$ and $\phi_{1}$. Therefore, an integrated energy
deposition at a given point (target) cannot be strictly monoenergetic,
even if the bulk of the beam is. Indeed, the head of the beam (which
is at $\psi=0)$ escapes the bulk of it with the speed $2\sqrt{\phi_{1}}$
(one may use eq.(\ref{eq:fi1} for $\phi_{1}$). However, the density
of these fast moving beam particles is nominally zero, while the bulk
of the beam has the density $\rho_{1}$, eq.(\ref{eq:rho1}). Therefore,
the net effect of this beam energy spreading needs to be investigated
depending on the nature of the target. Such investigation is beyond
the scope of the present paper. We merely mention here that from the
perspective of the proton/carbon radiation therapy, for example, the
beam energy deposition is largely a collective phenomenon (e.g., \cite{2015PhT....68j..28P}
and referenced therein). If so, then the beam energy density $\rho V_{b}^{2}$/2
is probably more relevant than the individual particle energy, $m_{i}V_{b}^{2}/2.$
Therefore, dumping the rarefied head will not necessarily result in
a significant additional spreading of the ``hot spot'' produced
by the bulk of the beam.

Notwithstanding the above remarks, it is worthwhile to calculate the
velocity spread of the beam. For cold upstream ions, we may neglect
this spread for the bulk of the beam that carries the potential $\psi=\phi_{1}$,
and calculate the spread for its head, where $0<\psi<\phi_{1}$, using
the approximation, $\phi_{1}\ll w^{2}$. Defining the beam velocity
spread as

\[
\Delta u=\frac{\int u\rho du}{\int\rho du}
\]
where $0<u<u_{{\rm max}}\simeq2\sqrt{\phi_{1}}$, using eqs.(\ref{eq:EOS})
and (\ref{eq:uOfPsi}), we obtain for $\Delta u$ the following simple
result

\[
\Delta u=u_{{\rm max}}/4\ll V_{b}
\]
The beam density $\rho$, is falling off with its velocity as follows

\[
\rho\left(u\right)=\frac{1}{4}\left(1-w^{-2}\right)\left(u_{{\rm max}}-u\right)^{2}.
\]
One sees that the beam velocity distribution remains relatively narrow
despite the acceleration of particles from its front. Also, the relative
contribution to the integrated energy deposition of the head of the
beam can be reduced by increasing the length of the primary beam;
that is the system length.

\section{Discussion and Conclusions\label{sec:Discussion}}

A better understanding of ion-acoustic collisionless shocks, including
ion reflection, is required for the operation of laser-based accelerators
\cite{BulanovRMP06,Dudnikova11PRL,Haberberger2012NatPh,FiuzaUCLA2013,Macchi13RMP}
(and many other applications, mentioned in passing in the Introduction
section). Turning to the astrophysical applications, by far the most
demanded particle acceleration mechanism, the diffusive shock acceleration
(DSA) is also likely to be fed in by the shock-reflected particles.
Although the DSA operates in magnetized plasmas, typically at much
larger than Debye scale, the particle reflection can hardly be understood
without understanding the DSA microscopics, to which the results of
the present paper are directly relevant. Identifying a seed population
(\textquotedblleft injected\textquotedblright{} particles) for the
DSA in the background plasma and understanding their selection mechanisms
\cite{mv95,m98,ZankInj01} presents a genuine challenge for interpreting
the new, unprecedentedly accurate observations of cosmic rays, e.g.,
\cite{Adriani11,AMS02_2014}. These observations point to the elemental
discrimination of particle acceleration that almost certainly is a
carry-over from the injection of thermal particles into the DSA \cite{MDSPamela12}.
Operating at the outer shocks of the supernova remnants, the DSA is
the basis of contemporary models for the origin of galactic cosmic
rays \cite{BerezBook90,Hillas05,Drury12,Gaisser2013,BlandfordCRorig2014,2014BrJPh..44..415B}.

Injection has been studied numerically mostly with hybrid simulations
\cite{KuchScholer91,Scholer02,Burgess12,Giacalone2013,CaprioliInj15}.
An accurate calculation of injection efficiency using the results
of the present paper would go far beyond its scope and focus. At a
minimum, such calculation must include the magnetic shock structure.
Conversely, the particle reflection analyzes for magnetized shocks
presented in many publications, e.g. \cite{Woods71,Gedalin08,Zank96,LeeSS96},
as well as the above-cited hybrid simulations, do not include the
electrostatic structure into the reflection process self-consistently
with electron and ion kinetics. In this paper, we addressed the questions
of how does the reflection affect the shock speed, its structure and
reflected ions themselves. We have determined their distribution,
given that of the incident ions and the shock Mach number. These results
will, therefore, be important for the comprehensive DSA injection
models yet to be build. Note that in the case of magnetized quasi-parallel
shocks, the injection seed particles other than reflected ones have
also been considered (see, e.g., \cite{Giacalone2013} for a recent
discussion of the alternatives). In particular, the thermalized downstream
particles have long been deemed to be a viable source for injection
\cite{EdmistonKennelEichler82} (so-called thermal leakage). One may
argue, however, that if such leakage occurs from the downstream region
within 1-2 Larmor radii off the shock ramp, the difference between
them and reflected particles is rather semantic from the DSA perspective
\cite{MDSPamela12}. 

We further highlight the following findings of this paper: (i) when
the soliton Mach number increases to the point of ion reflection,
and the soliton transforms into a soliton train downstream, this structure
persists with the increasing Mach number until most of the incident
ions reflect off the first soliton \footnote{In fact, almost all of them, when $T_{i}/T_{e}\to0$.}.
The reflection coefficient approaches $\alpha=\alpha_{c}\simeq1-3.9\sqrt{T_{i}/T_{e}}$,
(ii) at this point the downstream potential is equal to $\phi_{max}\simeq M_{*}^{2}/2\simeq1.26$.
In addition, the foot rises to $\phi_{1}\simeq\ln\left(1+\alpha_{c}\right)/\left(1-1/8\phi_{{\rm max}}\right)\simeq0.77$
(for $\alpha_{c}\to1$) , so that the total shock Mach number approaches
$M=M_{2}\simeq1.8$. This result is obtained for the Boltzmannian
electrons, while in the case of adiabatically trapped electrons the
maximum Mach number approaches $M=M_{2}\simeq4.5$, (iii) the laminar
shock structure cannot continue beyond this point. 

Based on the numerous PIC simulations, available in the literature
(e.g., \cite{Haberberger2012NatPh,FiuzaUCLA2013,Liseykina15}), we
may speculate that when the Mach number exceeds its critical value
$M_{2},$ obtained in this paper, the shock evolution becomes time
dependent; ions reflect intermittently. One example of such dynamics,
Fig.\ref{fig:Simulations}, we adopted from the recent PIC simulations
\cite{Liseykina15} (see also Appendix for a further brief discussion
of this result). For yet higher Mach numbers, the upstream and downstream
flows do not couple together, but rather penetrate through each other,
not being perturbed significantly. In a piston driven flow, ions reflect
only from the piston, so the shock does not form. As for the prospects
for a laser-based accelerator, this is probably a favorable scenario
for generating ion beams when the high energy is a priority. Indeed,
the maximum Mach number for a laminar shock, with sustainable ion
reflection from its front, is rather low. Therefore, ions reflected
directly from the piston may be a better solution. 

In this paper, the main part of the shock structure was resolved exactly,
by adopting a simplified kinetic model for a finite-temperature \textquotedblleft box\textquotedblright{}
distribution of upstream ions, using the shock pseudopotential. Considering
$T_{i}/T_{e}$ as a small parameter, the number of reflected ions
is calculated as a function of the shock Mach number self-consistently
with the shock foot potential. The dynamics of the reflected ion beam
in the foot is investigated.

To recapitulate the relation of this and earlier studies, we note
that many analyzes were limited to the case of monoenergetic upstream
ions. For that reason, they could not resolve ion reflection as the
incident ions should all reflect at once, when the peak of wave potential
$e\phi_{max}$ becomes equal to the ion energy $m_{i}V_{{\rm shock}}^{2}/2$.
As we pointed out already, this happens when the Mach number $M=\sqrt{V_{s}/C_{s}}$
reaches $M=M_{*}\approx1.6$ for Boltzmann electrons \cite{Sagdeev66},
and $M=M_{*}\approx3.1$, for adiabatically trapped electrons \cite{Gurevich68}.
Numerical treatments, however, included finite ion temperature and
have been able to address the effect of ion reflection on the shock
structure by using PIC simulations, e.g. \cite{FiuzaUCLA2013,Liseykina15}.
The ion reflection alters the shock amplitude and speed, thus impacting
the reflection threshold itself. The most striking result of this
feedback loop, we have studied in this paper, is a pedestal of the
electrostatic potential, built upstream. It changes the speed of inflowing
ions and thus, again, the condition for their subsequent reflection
from the main shock. To our best knowledge, this important aspect
of the collisionless shock physics has not yet been studied systematically
in PIC simulations.
\begin{acknowledgments}
MM would like to thank University of Maryland for the hospitality
and support during this work. He and PD also acknoledge support from
NASA ATP-program under grant NNX14AH36G and Department of Energy under
Award No. DE-FG02- 04ER54738.
\end{acknowledgments}

\appendix*

\section*{Appendix}

\setcounter{equation}{0}

By analogy with their gasdynamics counterparts, we rewrite eqs.(\ref{eq:ContPrec}-\ref{eq:MomenPrec})
in a form of two Riemann invariants conserved along two families of
characteristics. To this end, we first change the dependent variable
in eq.(\ref{eq:ContPrec}) $\rho\mapsto J$, so that this equation
rewrites: 

\begin{equation}
\frac{\partial J}{\partial t}+u\frac{\partial J}{\partial x}+\sqrt{\rho\left(\frac{\partial\rho}{\partial\psi}\right)^{-1}}\frac{\partial u}{\partial x}=0\label{eq:Jeq}
\end{equation}
where

\begin{equation}
J=\intop d\psi\sqrt{\frac{\partial\ln\rho}{\partial\psi}}\label{eq:Jdef}
\end{equation}
By summing and negating Eqs.(\ref{eq:MomenPrec}) and (\ref{eq:Jeq}),
we arrive at the following characteristic form of them

\begin{equation}
\frac{\partial R_{\pm}}{\partial t}+C_{\pm}\frac{\partial R_{\pm}}{\partial x}=0\label{eq:RiemannCons}
\end{equation}
with the Riemann's invariants $R_{\pm}$ and the characteristics $C_{\pm}$,
respectively, given by

\begin{equation}
R_{\pm}=u\pm\intop_{\phi_{1}}^{\psi}d\psi\sqrt{\frac{\partial\ln\rho}{\partial\psi}}\;\;\;{\rm and}\;\;\;C_{\pm}=u\pm\left(\frac{\partial\ln\rho}{\partial\psi}\right)^{-1/2}.\label{eq:RiemDef}
\end{equation}
Therefore, the most general solution of the problem, described in
Sec.\ref{sec:SolutionPrecursor} by eqs.(\ref{eq:ContPrec}-\ref{eq:MomenPrec}),
is determined by conservation of $R_{\pm}$ along the characteristics
$C_{\pm}$. From this perspective, the simple wave solution given
by eqs.(\ref{eq:PsiSol}-\ref{eq:ksi2}) corresponds a decaying discontinuity
with $u\left(x<0\right)=0$ and $u\left(x\ge0\right)=u_{1}\equiv u\left(\psi=0\right)$,
(eq.{[}\ref{eq:uOfPsi}{]}). The initial beam density jump is defined
in a similar way, $\rho\left(x\geq0\right)=0$ and $\rho\left(x<0\right)=\rho_{1}$,
eq.(\ref{eq:rho1}). Under these initial conditions, the Riemann's
invariant $R_{+}\equiv0$ everywhere. Thus, the initial value problem
given by eq.(\ref{eq:RiemannCons}), significantly simplifies with
only $R_{-}\neq0$ and a single family of characteristics $C_{-}$
involved in it. As $C_{-}$ characteristics diverge from the origin,
the simple wave solution described in Sec.\ref{sec:SolutionPrecursor}
emerges, and it is consistent with the initial conditions specified
above. 

\begin{figure}
\includegraphics{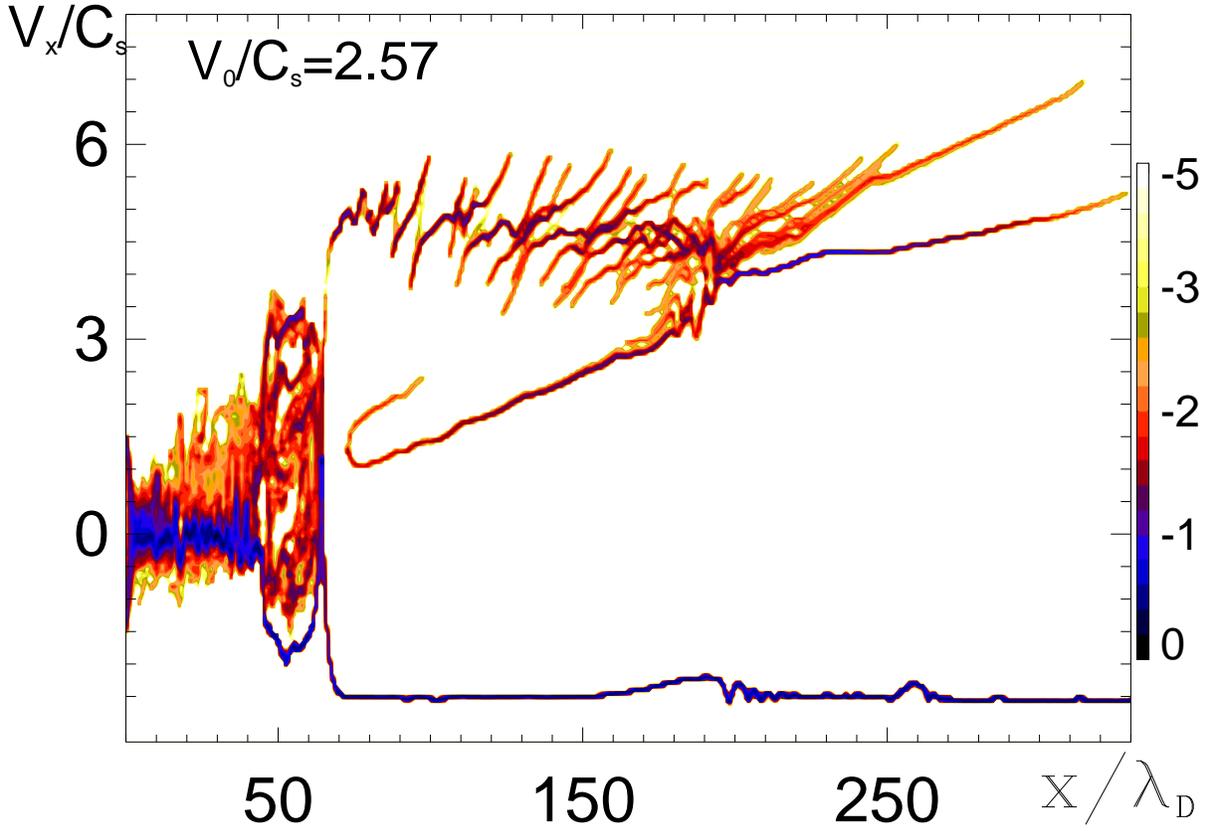}

\caption{PIC simulation result from Ref. \cite{Liseykina15}. Shown is the
ion phase plane at $t\omega_{pe}=2800$, $M_{0}=V_{0}/C_{s}=2.57$;
the resulting velocity of the shock is around 3.8. The color coding
corresponds to the ion phase space density normalized to that of the
upstream ions (logarithmic scale).\label{fig:Simulations}}
\end{figure}

It is important to emphasize that under more general initial conditions,
the beam dynamics can be much more complicated. In particular, the
flow characteristics generally intersect. As the beam ``hydrodynamics''
is truly collisionless, their intersection will result in a multi-flow
state of the reflected beam. Such states copiously emerge in simulations,
e.g., \cite{Macchi13,FiuzaUCLA2013,Liseykina15}, along with laminar
reflected beam flows described in Sec.\ref{sec:SolutionPrecursor}.
An illustrative example, taken from recent PIC simulations \cite{Liseykina15},
is shown in Fig.\ref{fig:Simulations}. Even though the shock is super-critical,
the quasi-laminar part of the reflected ion beam, described in this
paper, can be easily identified in the area $x>175$. Here the flat
part of beam density distribution ($175<x<200)$ transitions into
an accelerating, rarefied part at $x>200$. Other reflected ion components
in this area stem from later, non-stationary and highly intermittent
reflection events. Being more energetic, these ions are catching up
with the laminar part at the moment shown in the Figure. Based on
the color coding, however, they are considerably (about 10 times)
lower in phase space density than the main reflected component is.

\bibliographystyle{C:/Users/mmalkov/bibtex/prsty}
\bibliography{C:/Users/mmalkov/bibtex/textbooks,C:/Users/mmalkov/bibtex/PlasmaDSA,C:/Users/mmalkov/bibtex/LaserAccel,C:/Users/mmalkov/bibtex/MALKOV,C:/Users/mmalkov/bibtex/dsa,C:/Users/mmalkov/bibtex/SNR,C:/Users/mmalkov/bibtex/DSAobs,C:/Users/mmalkov/bibtex/SolarWindHeliosph}

\end{document}